\let\pdfoutput\defined
\documentclass{PoS}
\def\lsim{\raise0.3ex\hbox{$<$\kern-0.75em\raise-1.1ex\hbox{$\sim$}}}
\def\gsim{\raise0.3ex\hbox{$>$\kern-0.75em\raise-1.1ex\hbox{$\sim$}}}
\title{%
\vspace*{-1cm}
\begin{minipage}{\textwidth}
\begin{flushright}
\texttt{\footnotesize
PoS(LAT2006)021\\%
BNL-NT-06/40\\%
}
\end{flushright}
\end{minipage}\\[15pt]
Lattice QCD at finite density}
\ShortTitle{Lattice QCD at Finite Density}

\author{\speaker{Christian Schmidt}\\
        Brookhaven National Laboratory, Physics Department, Upton, NY 11955, USA\\
        E-mail: \email{cschmidt@bnl.gov}}

\abstract{I discuss different approaches to finite density lattice QCD. $\;$In particular, I focus on the structure of the phase diagram and discuss 
attempts to determine the location of the critical end-point. Recent results 
on the transiton line as function of the chemical potential ($T_c(\mu_q)$) 
are reviewed. Along the transition line, hadronic fluctuations have been 
calculated, which can be used to characterize properties of the Quark Gluon 
plasma and eventually can also help to identify the location of the critical 
end-point in the QCD phase diagram on the lattice and in heavy ion
experiments. Furthermore, I comment on the structure of the phase diagram at
large $\mu_q$.

}

\FullConference{XXIVth International Symposium on Lattice Field Theory\\
                July 23-28, 2006\\
                Tucson, Arizona, USA}

\begin{document}
\section{Introduction}
\begin{figure}
\begin{center}
\includegraphics[width=.4\textwidth]{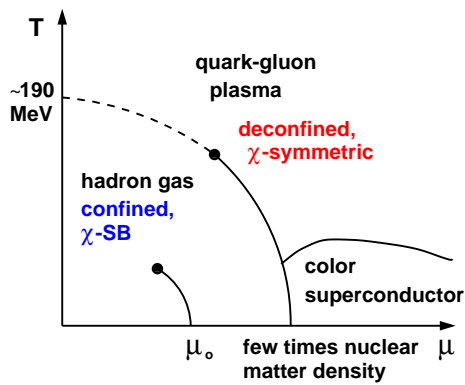}
\end{center}
\caption{Sketch of the QCD phase diagram.}
\label{phasediagram}
\end{figure}
Lattice QCD currently is the only quantitative approach to finite temperature
QCD based on first principle calculation. For a recent review see 
\cite{Heller:2006ub}. At non zero density however, lattice 
QCD is harmed by the sign problem ever since its inception. The Fermion
matrix becomes complex and can not be interpreted as a probability 
distribution. Hence straight forward Monte Carlo simulations become impossible.
For a detailed description of the sign problem in the epsilon regime see 
\cite{Kim}.

During the last few years a lot of progress has been made to circumvent
the sign problem for small values of $\mu_q/T$, where $\mu_q$ is the
quark chemical potential and $T$ the temperature. This progress helps to 
understand the physics relevant for heavy ion collisions and
eventually will clarify the existence/location of the critical 
end-point in the QCD phase diagram. In Fig.~\ref{phasediagram} a sketch 
of the QCD phase diagram in the $T$-$\mu$ plane is shown. Lattice QCD 
calculations provide more and more evidence that the QCD transition 
at $\mu_q=0$ is not a phase transition in the thermodynamic sense, but a smooth
crossover. Further evidence was seen recently in \cite{Wuppertal}.
Nevertheless, one can define a transition temperature $T_c$
by the peak position of the chiral susceptibility. As a function of the 
quark chemical potential the line of transition temperatures 
($T_c(\mu_q)$) is smoothly connected to a critical end-point in the 
($T,\mu_q$)-diagram. For larger chemical potentials the QCD transition is 
expected to be first order. At high densities, several color 
superconducting phases are expected.   

The rest of the article is organized as follows: in Sec.~\ref{sec:methods}
I will briefly recall different methods which have been used so far 
to calculate thermodynamic observables at non zero chemical potential.
In Sec.~\ref{sec:line} I will summarize current knowledge about the 
$\mu_q$-dependence of the critical temperature. I will continue with 
reviewing results on quark number fluctuations along the transition
line (Sec.~\ref{sec:fluct}) and the critical point (Sec.~\ref{sec:EP}).
Finally I will discuss the physics beyond the critical point in 
Sec.~\ref{sec:beyond}.

\section{Methods to extract information on the chemical potential 
dependence}
\label{sec:methods}

\subsection{Reweighting from the ($\mu=0$)-ensemble}
\label{sec:rew}
On the lattice one has to choose several parameters to characterize a
thermodynamic system. In addition to the number of lattice points in 
spacial and temporal directions, $N_s, N_t$ respectively, we have to 
choose quark masses $m_q$, the coupling $\beta\equiv 6/g^2$ and the 
chemical potential $\mu_q$. Together these parameters define the lattice 
spacing $a$ and thus also the temperate $T\equiv 1/aN_t$ and volume 
$V\equiv (a N_s)^3$ of the simulated system. A thermodynamic observable 
is calculated on the lattice as   
\begin{equation}
\left<O\right>_{\beta,m_q,\mu_q}=\frac{1}{Z(V,T)}\int {\mathcal{D}U}\;
O\;\left[{\rm det}M(U;m_q,\mu_q)\right]^{N_f/4}\; \exp\{-\beta S_G(U)\} 
\quad . \label{EqO}
\end{equation}
Here $N_f$ is the number of dynamical fermions. The notation is written 
down for staggered fermions as the additional factor of $1/4$ in the 
power of the fermion determinant indicates. See \cite{Sharpe} for a 
discussion of the ``4th root trick'' needed to calculate the staggered
fermion determinant.

In principle it is possible to calculate the expectation value of the 
observable at the parameter set $p=\{\beta,m_q,\mu_q\}$, from an ensemble
generated at $p_0=\{\beta_0,m_{q,0},\mu_{q,0}\}$. We have the identity 
\begin{equation}
\left<O\right>_p=\left<O\;R(U;p,p_0)\right>_{p_0} / 
\left< R(U;p,p_0)\right>_{p_0} \quad ,
\end{equation}
where we define the reweighting factor $R$ as
\begin{equation}
R(U;p,p_0)\equiv \left[{\rm det}M(U;p)/{\rm det}M(U;p_0)\right]^{N_f/4}
\;\exp\{-(\beta-\beta_0)S_G\} \quad .
\label{EqR}
\end{equation}
The reweighting method as a tool to perform extrapolation and interpolation 
in the gauge coupling $\beta$ goes back to \cite{FS}. For reweighting in the 
chemical potential it was first used by the Glasgow group \cite{Glasgow}. 
However, since the overlap between the generated ensemble at $\mu_{q,0}=0$ 
and the target ensemble at $\mu_q>0$ exponentially decreases with increasing 
$\mu_q$, the method was successful only after it was generalized to a 
multi-parameter approach \cite{FKmethod}.
For $N_t=4$ lattices it was found that reweighting along the transition 
line $T_c(\mu_q)$ works quite well up to $a\mu_q\lsim 0.3$ or equivalently
for $\mu_q/T\lsim 1.2$.

In general the reweighting approach requires the evaluation of the
fermion determinant on every configuration. As this is computationally 
demanding, one may consider to expand the reweighting factor given in
Eq.~(\ref{EqR}) in terms of the chemical potential \cite{BiSw1}. In this 
case the reweighting procedure is, however, only correct up to a certain 
order in $\mu_q/T$. 

\subsection{The Taylor expansion method}
It is conceptually very simple to calculate the expansion coefficients
of any observable $O$ (Eq.~(\ref{EqO})) in a Taylor series around $\mu_q=0$:
\begin{equation}
O(\hat{\mu})=c_0+c_1\hat{\mu}+\frac{1}{2}c_2\hat{\mu}^2+\cdots \quad .
\label{taylor_series_O}
\end{equation}
Since on the lattice all quantities are given in units of the lattice spacing
($a$), the expansion parameter is $\hat{\mu}\equiv a\mu_q=N_t^{-1}(\mu_q/T)$. 
This idea goes back to the first calculation of the quark number susceptibility
\cite{Gottlieb:1988cq}. The response of meson masses \cite{Choe:2001ar} as 
well as the pressure and further bulk thermodynamic quantities 
\cite{Gavai:2003mf,Allton:2003vx,Allton:2005gk,Ejiri:2005uv} have been studied 
by this method. The first two nontrivial coefficients in 
Eq.~(\ref{taylor_series_O}) are given by
\begin{eqnarray}\label{taylor_coeff_O}
c_1&=&  \left< \frac{\partial O           }{\partial \hat{\mu}}\right>
      \!+\! \left<O\frac{\partial \ln\rm{det}M}{\partial \hat{\mu}}\right>\\
c_2&=&  \left< \frac{\partial^2 O}{\partial \hat{\mu}^2}\right>
      \!+\!2\left< \frac{\partial O}{\partial \hat{\mu}}
               \frac{\partial \ln\rm{det}M}{\partial \hat{\mu}}\right>
      \!+\! \left<O\frac{\partial^2 \ln\rm{det}M}{\partial \hat{\mu}^2}\right>
      \!-\! \bigg<O\bigg>\left<\frac{\partial^2 \ln\rm{det}M}{\partial
      \hat{\mu}^2}\right> \quad . \nonumber
\end{eqnarray}
Besides derivatives of the observable itself, the calculation of
derivatives of $\ln \rm{det}M$ with respect to $\hat{\mu}$ is required.
The derivatives have to be taken at $\hat{\mu}_0=0$. Note that due to a
symmetry of the partition function ($Z(\mu_q)=Z(-\mu_q))$ all odd 
coefficients in Eq.~(\ref{taylor_series_O})  vanish identically. For the 
same reason we have
$\left<\partial \ln\rm{det} M/\partial \hat{\mu}\right>=0$ at
$\hat{\mu}=0$. We explicitly use this property in
Eq.~(\ref{taylor_coeff_O}) to derive the expansion coefficients. 

The advantages of this method are that expectations values only have to be
evaluated at $\hat{\mu}=0$, i.e. calculations are not directly affected by 
the sign problem.  Furthermore, all derivatives of the fermion determinant 
can be expressed in terms of traces by using the identity 
$\ln\rm{det}M=\rm{Tr}\ln M$. This enables the stochastic calculation of the 
expansion coefficients by the random noise method, which is much faster than 
a direct evaluation of the determinant. Moreover, the continuum and infinite 
volume extrapolations are well defined on a coefficient by coefficient basis.

On the other hand it is a priori not clear for how large $\mu/T$ the method
works and how large the truncation errors are. Furthermore one is strictly
limited by phase transitions, since phase transitions are connected with
discontinuities or divergences. An estimation of the convergence radius of
the series gives a lower bound on the applicability range and thus also a
lower bound to the phase transition line in the $(T,\mu)$ plane (see the
discussion in Sec.~\ref{sec:EP}).

\subsection{Analytic continuation}
\label{sec:imag}
At imaginary chemical potentials, the fermion determinant is real and
positive, thus simulations by standard Monte Carlo techniques are
pos\-sible. Results on the imaginary $\hat\mu_I$ axis can be analytically
continued to the real $\hat\mu_R$ axis. It is especially easy to convert a
Taylor series in $\hat\mu_I$, expanded around $\hat\mu=0$, into a Taylor
series in $\hat\mu_R$. Since the series has only even powers of $\hat\mu$, due
to the the symmetry $Z(\hat\mu)=Z(-\hat\mu)$, one only has to switch the sign
of every second coefficient ($c_2\rightarrow-c_2, c_6\rightarrow-c_6,
\dots$). There is however another symmetry of the partition function which
limits the analytic continuation. Due to the periodicity \cite{Roberge:1986mm}
$Z(\mu_R,\mu_I)=Z(\mu_R,\mu_I+2\pi T/3)$ simulations with $\mu_I>0$ will only
give access to the physical region $\mu_R\lsim \pi T/3$. This method was used 
to map out the phase transition line 
\cite{deForcrand:2003hx}-\cite{D'Elia:2002gd}. One should note,
that for this method neither an evaluation of the determinant, nor any of its
derivatives is required. In order to determine the functional dependence of an
observable on $\hat\mu_R$ one needs many different simulation points for 
several values of $\hat\mu_I$, to perform an analytic continuation using a 
certain Ansatz.

\begin{figure}
\begin{center}
\begin{minipage}[b]{.48\textwidth}
\includegraphics[width=.84\textwidth, angle=-90]{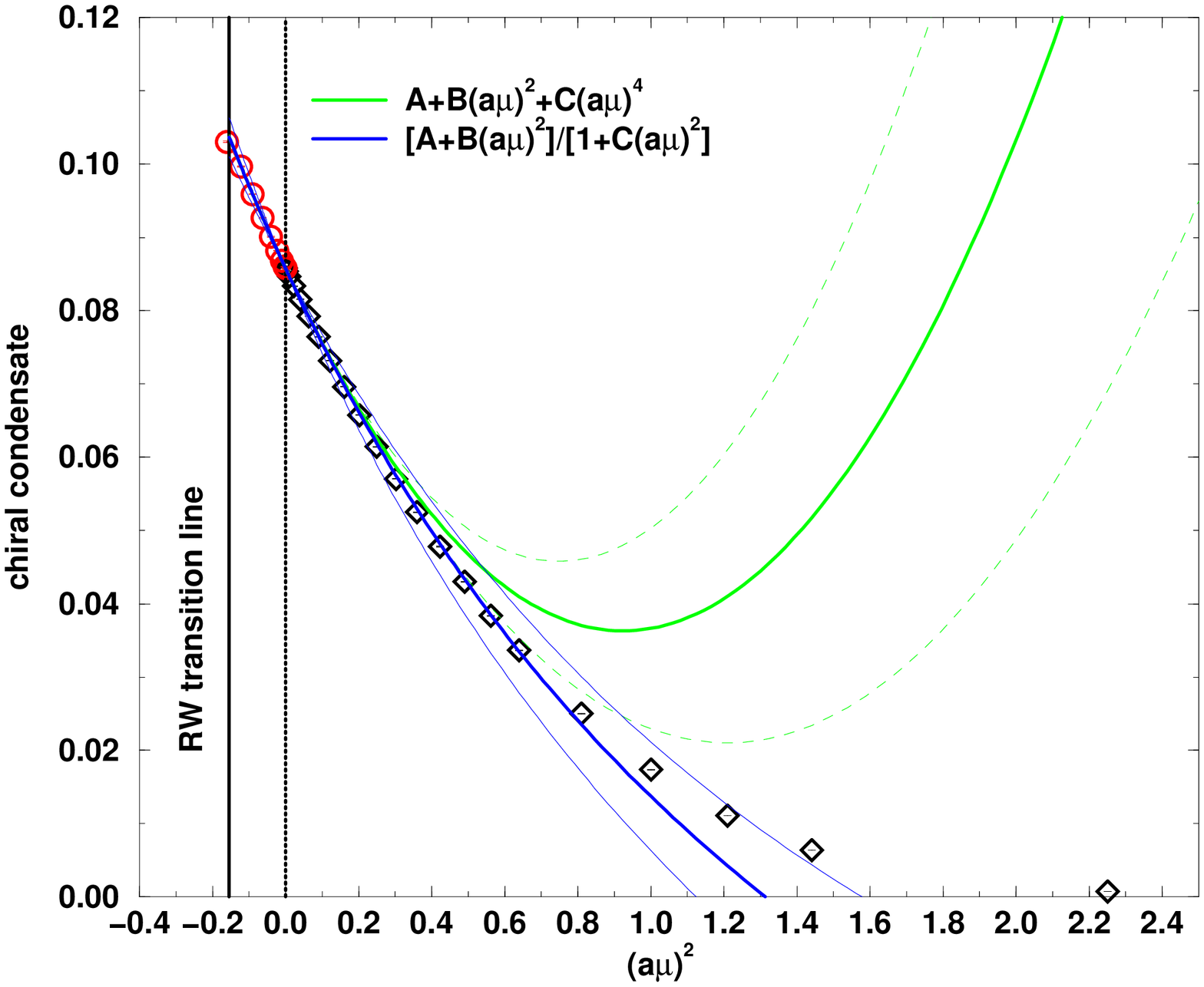}
\caption{The chiral condensate as a function of $\hat\mu^2$, in $SU(2)$ gauge
theory [19]. The solid lines are fits with different Ans\"atze to the data for  
$\hat\mu^2<0$. \label{fig:imag1}}
\end{minipage}\hfill
\begin{minipage}[b]{.48\textwidth}
\includegraphics[width=.7\textwidth, angle=-90]{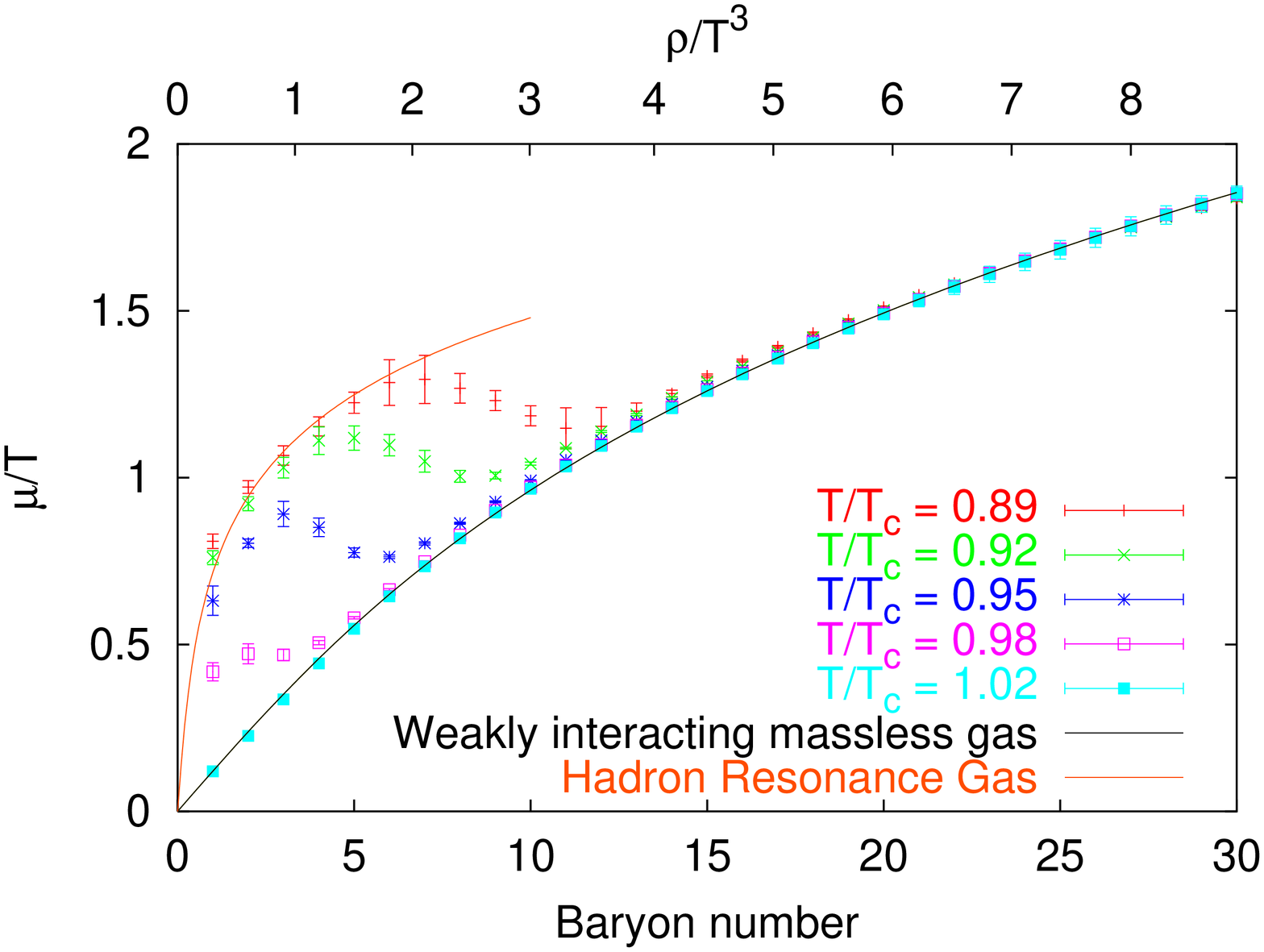}
\caption{Relation between the Baryon number and the chemical potential from 
the saddle point approximation, for different temperatures [27].
\label{fig:saddle}}
\end{minipage}
\end{center}
\end{figure}

A demonstration of the imaginary chemical potential method is given in
Fig.~\ref{fig:imag1}. 
The method was applied to the case of 2-color QCD 
\cite{Papa}. Here it is not only possible to calculate observables at
$\hat\mu^2<0$, but also at $\hat\mu^2>0$, since 2-color QCD does not suffer
from a sign problem. Thus one can explicitly check how far the extrapolation
in $\hat\mu^2$ is valid and which type of Ansatz is especially well suited for
the extrapolation. In general, extrapolations with rational functions seem to 
be better than extrapolations with polynomials \cite{Papa, Pade}.

\subsection{The canonical approach}
\label{sec:canonical}
The canonical partition function ($Z_C$) can be constructed by introducing a 
$\delta$-function into the grand canonical partition function which fixes
the net number of quarks present in the system. Using an integral 
representation of the $\delta$-function one recognizes the integration 
parameter as an imaginary chemical potential. One finds
\begin{equation}
Z_C(T,Q=3B)=\frac{1}{2\pi}\int_{-\pi}^{\pi} {\rm d}
\left(\frac{\mu_I}{T}\right)\;
\exp\left\{-i3B\frac{\mu_I}{T}\right\}Z_{GC}(T,i\mu_I)\quad.
\end{equation}
Thus, the canonical partition functions are the coefficients of the Fourier
expansion of the grand canonical partition function ($Z_{CG}$) in imaginary 
$\mu$. Here we have used the $2\pi T/3$ periodicity of the 
grand canonical partition function which as a consequence leads to the fact 
that the canonical partition functions vanish for non-integer values of the 
baryon number $B=3Q$.

The Fourier-coefficients can be computed exactly \cite{Hasenfratz:1991ax}. 
As for the reweighting method (\ref{sec:rew}), the evaluation of the 
fermion determinant is required on every configuration. In fact the same 
method can be used, which is the calculation of all $6N_s^3$ eigenvalues of 
the so-called ``reduced matrix'' \cite{FKmethod}.

After having calculated the canonical partition functions, a relation between 
the chemical potential and the baryon number is needed, in order to explore 
the phase diagram in the $(T,\mu)$-plane. Such a relation can be obtained by 
using the saddle point approximation of the fugacity expansion (which is 
exact in the  thermodynamic limit). One finds
\begin{equation}
\mu(\rho)=\frac{1}{3}\frac{\partial f(\rho)}{\partial\rho}\quad,
\end{equation} 
where $\rho=B/V$ is the baryon number density and 
$f(T,\rho)=-\frac{T}{V}{\rm log}Z_C(T,\rho)$ is the Helmhotz free energy 
density. For several different temperatures the saddle point approximation is 
shown in Fig.~\ref{fig:saddle} \cite{canonical}. Due to the computational 
costs the calculation have been performed on $6^3\times 4$ lattices, with 
$N_f=4$ flavors of staggered fermions. As can be seen from the ``S''-shape of 
the curves, one can have more than one solution when solving for the baryon 
density $\rho$, at given $\mu$ and $T$. This reflects the nature of the 
transition in the four-flavor theory, which is of first order. By using a 
Maxwell construction, one is able to calculate the two densities 
$\rho_1(T)<\rho_2(T)$, giving the lower and upper boundary of the 
coexistence area, as well as the critical chemical potential 
$\mu_B^{crit}(T)$, as functions of the temperature. 

\subsection{The density of states method}
\label{sec:dos}
An alternative to the importance sampling technique used in most Monte Carlo
simulations is the density of states method. Here one reorders the path
integral representation of the partition function in the following way: first 
expectation values with a constrained parameter will be calculated. 
{\it I.e.} one selected parameter ($\phi$) is fixed. Expectation values 
according to the usual grand canonical partition function ($Z_{GC}$) can 
then be recovered by the integral
\begin{equation}
<O>=\int d\phi \, \left<Of(U)\right>_\phi \rho(\phi)
\left/ \int d\phi \, \left<f(U)\right>_\phi \rho(\phi)\right.
\end{equation}
where the density of states ($\rho$) is given by the constrained partition
function:
\begin{equation}
\rho(x)\equiv Z_\phi(x)=\int \mathcal{D}U\, g(U) \, \delta( \phi - x ) \quad .
\label{eq:dos}
\end{equation}
Here $\left<~\right>_\phi$ denotes the expectation value with respect to the
constrained partition function. In addition, the product of the weight
functions $f,g$ has to equal the correct measure of $Z_{GC}$:
$fg=\rm{det}M\cdot\exp\{-S_G\}$. The idea of reordering the partition
functions is rather old and was used successfully for gauge theories 
\cite{GAUGE} and QED with dynamical fermions \cite{QED}. For QCD the 
parameter $\phi$ is usually chosen to be the plaquette \cite{LUO}: $\phi=P$. 
In \cite{Gocksch}, however, the DOS method was constructed for the complex 
phase ($\phi=\theta$). Within the random matrix model, the authors of 
\cite{Ambjorn} used the quark number density ($\phi=n_q$).

The advantages of this additional integration becomes clear, when choosing
$\phi=P$ and $g(U)=1$. In this case $\rho(\phi)$ is independent of all
simulation parameters. The observable can be calculated as a function of all
values of the lattice coupling $\beta$. If one has stored all eigenvalues of
the fermion matrix for all configurations, the observable can also be
calculated as a function of quark mass ($m$) and number of flavors\cite{LUO}
($N_f$).

Note that this method does not solve the sign problem. It is, however supposed
to solve the overlap problem. Moreover, it is also possible to combine the DOS
method, with the reweighting method \ref{sec:rew}, by reweighting the
constrained expectation values in the case of $g(U)\ne 1$. For large
reweighting distances an overlap problem is then introduced once again.

\section{The transition line}
\label{sec:line}
\begin{figure}
\begin{center}
\begin{minipage}[b]{.58\textwidth}
\includegraphics[width=.70\textwidth, angle=-90]{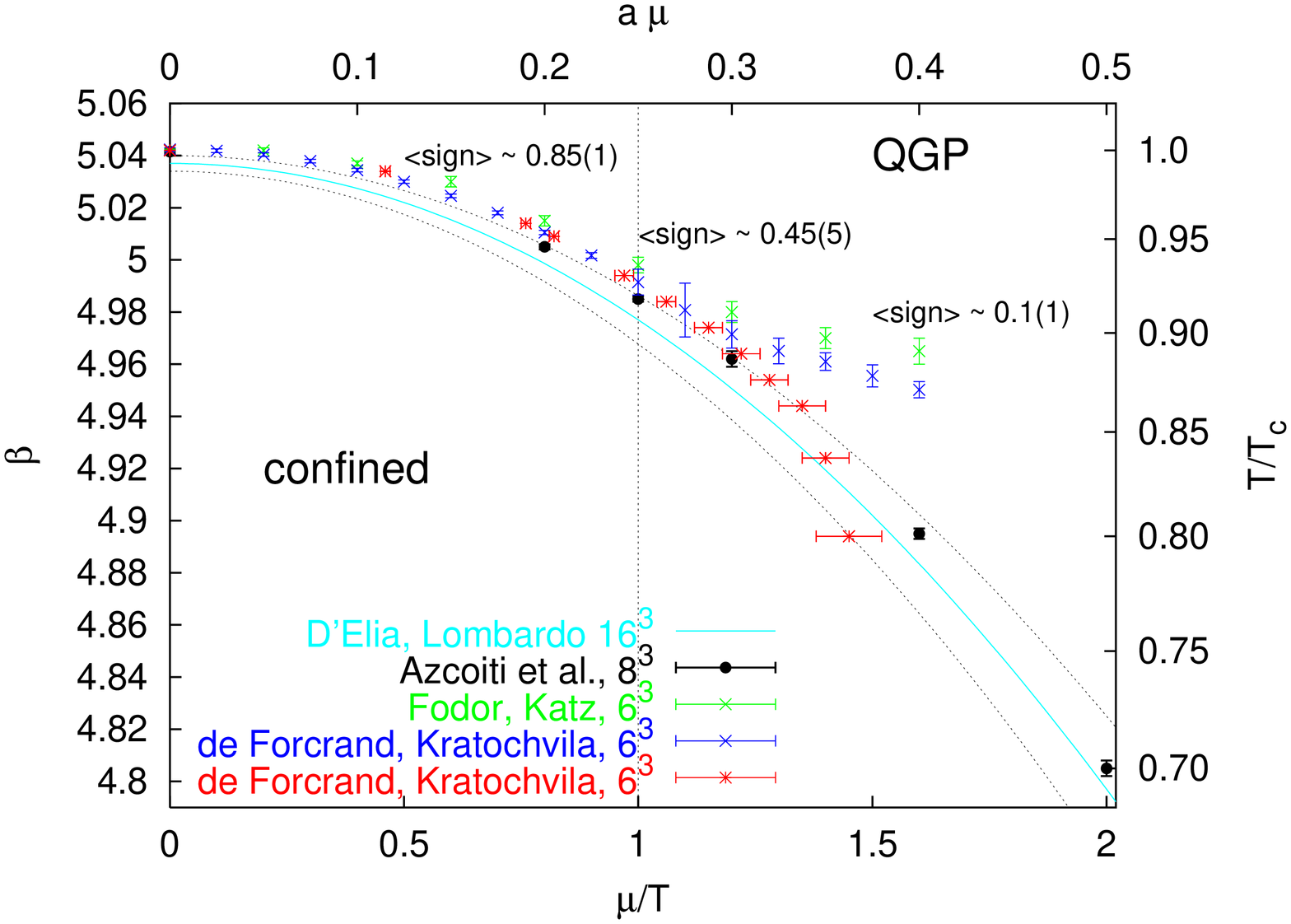}
\caption{Comparison of the transition line $T_c(\mu)$ obtained from different
methods. Plot from [27]. \label{fig:tccomp}} 
\end{minipage}\hfill
\begin{minipage}[b]{.38\textwidth}
\includegraphics[width=.99\textwidth]{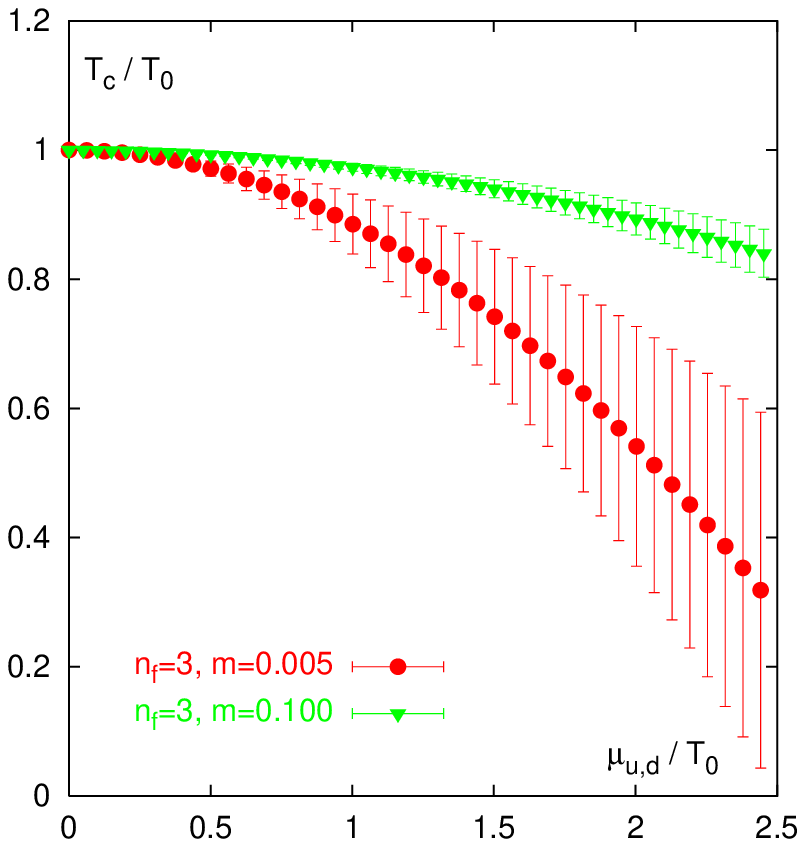}
\caption{The mass dependence of the transition line for
  $N_f=3$ [29]. \label{fig:tcmass}} 
\end{minipage}
\end{center}
\end{figure}

Using any of the methods presented above, the calculation of the transition 
line $T_c(\mu_q)$ is possible. This has been done for many systems,
which differ in the number of quark flavors, quark masses, physical volume 
and lattice spacing. This makes a comparison of different methods difficult.
The case of $N_f=4$, $m/T=0.2$ and $N_t=4$, however, has been studied
extensively with almost all methods. A comparison is given in
Fig.~\ref{fig:tccomp} \cite{canonical}. 
As one can see, the agreement between different methods is very good up to
$a\mu_q\approx 0.3$ or equivalently $\mu_q/T\approx 1.2$. For larger chemical
potential the two results from the reweighting method (Sec.~\ref{sec:rew}),
indicated as green and blue points, seem to stay above the other results.
The reason could be the lack of overlap between the simulated and the 
reweighted ensemble. 

Other results shown in Fig.~\ref{fig:tccomp} come 
from the imaginary chemical potential approach (Sec.~\ref{sec:imag}), solid
line, from a generalized imaginary chemical potential approach, black dots,
and from the canonical approach (Sec.~\ref{sec:canonical}), red points.
These results seem to be in agreement even for somewhat larger chemical
potentials. However, the transition line from the canonical
approach bends down at $\mu/T\approx 1.5$. Strong 
coupling calculations at $\beta=0$ \cite{Kawamoto:2005mq} show that this is 
indeed necessary in order to match with the correct strong coupling limit.

To discuss the transition line a bit more quantitatively, one can 
expand $T_c(\mu_q)$ in terms of the chemical potential
\begin{equation}
\frac{T_c(\mu_q)}{T_c(0)}=1-t_2(N_f,m_q)\left(\frac{\mu_q}{\pi T}\right)^2
+{\mathcal O}\left( \left( \frac{\mu_q}{\pi T}\right)^4 \right) \quad .
\end{equation}  
In general the first non trivial coefficient $t_2$ will depend on the
number of flavors and the quark masses as indicated above. Of course they will
also be sensitive to finite volume and cut-off effects. One can, however, hope 
that for large physical volumes and small lattice spacings, i.e. $N_s/N_t>4$ 
and $N_t>4$, those effects are small. A detailed comparison of $t_2$ is given 
in Tab.~\ref{tab:tccomp}.
\begin{table}
\begin{center}
\begin{tabular}{cccccccc}\hline \hline 
$N_f$&$am$&$N_s$&$t_2$&Action&$\beta$-Function&Method&Reference    \\ \hline 
2  &0.1   & 16      &0.69(35) &p4   &non-pert.   &Taylor+Rew.& [8] \\   
   &0.032 & 6,8     &0.500(54)&stag.&2-loop pert.&Imag.      & [17]\\   
3  &0.1   & 16      &0.247(59)&p4   &non-pert.   &Taylor+Rew.& [29]\\ 
   &0.026 & 8,12,16 &0.667(6) &stag.&2-loop pert.&Imag.      & [39]\\ 
   &0.005 & 16      &1.13(45) &p4   &non-pert.   &Taylor+Rew.& [29]\\ 
4  &0.05  & 16      &1.86(2)  &stag.&2-loop pert.&Imag.      & [18]\\ \hline
2+1&0.0092,0.25&6-12&0.284(9) &stag.&non-pert.   &Rew.       & [34]\\ 
\hline \hline
\end{tabular}
\caption{Comparison of the first nontrivial coefficient $t_2$ in the Taylor 
expansion of the transition line. All results have been obtained with
$N_t=4$. \label{tab:tccomp}}
\end{center}
\end{table}
In general, the curvature of the transition line becomes steeper for 
increasing number of flavors and decreasing quark masses. Two of the 3-flavor 
results which have been obtained with the p4-improved action 
\cite{Karsch:2003va} are shown in Fig.~\ref{fig:tcmass}. In some cases it has 
also been possible to estimate the sub-leading coefficient $t_4$, which has 
been found to be very small or even negative. If an
extrapolation with a Pad\'e Ansatz is performed, the transition
line tends to be steeper for high $\mu_q$ \cite{Pade} compared to the
truncated Taylor series and shows faster convergence.

We also note, that the calculation of $T_c(\mu_q)$ has two parts. The first 
part involves the calculation of $\beta_c(\hat\mu)$, the second one
is the calculation of the lattice $\beta$-function ($a\partial\beta/\partial
a$). Some of the results listed in Tab.~\ref{tab:tccomp} have been obtained by
using the perturbative two-loop $\beta$-function, which has the tendency to
underestimate the curvature of the critical line.

\section{Hadronic fluctuations}
\label{sec:fluct}
\begin{figure}
\begin{center}
\begin{minipage}[b]{0.48\textwidth}
\includegraphics[width=.95\textwidth]{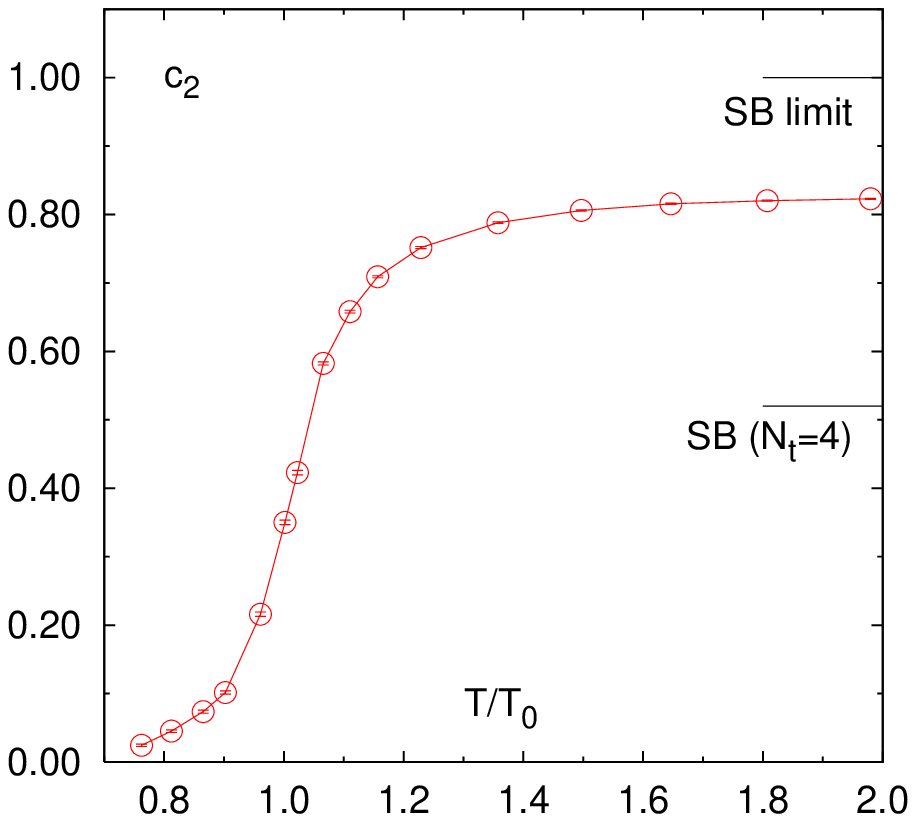}
\end{minipage}
\hfill
\begin{minipage}[b]{0.48\textwidth}
\includegraphics[width=.95\textwidth]{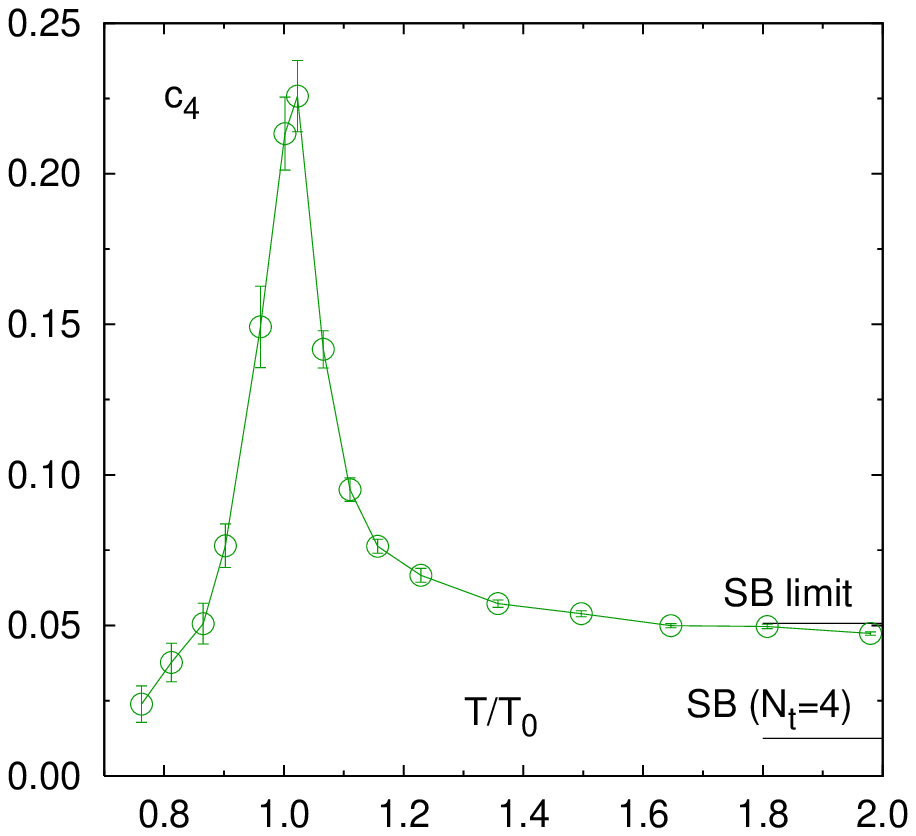}
\end{minipage}
\end{center}
\begin{minipage}[b]{0.48\textwidth}
\includegraphics[width=.95\textwidth]{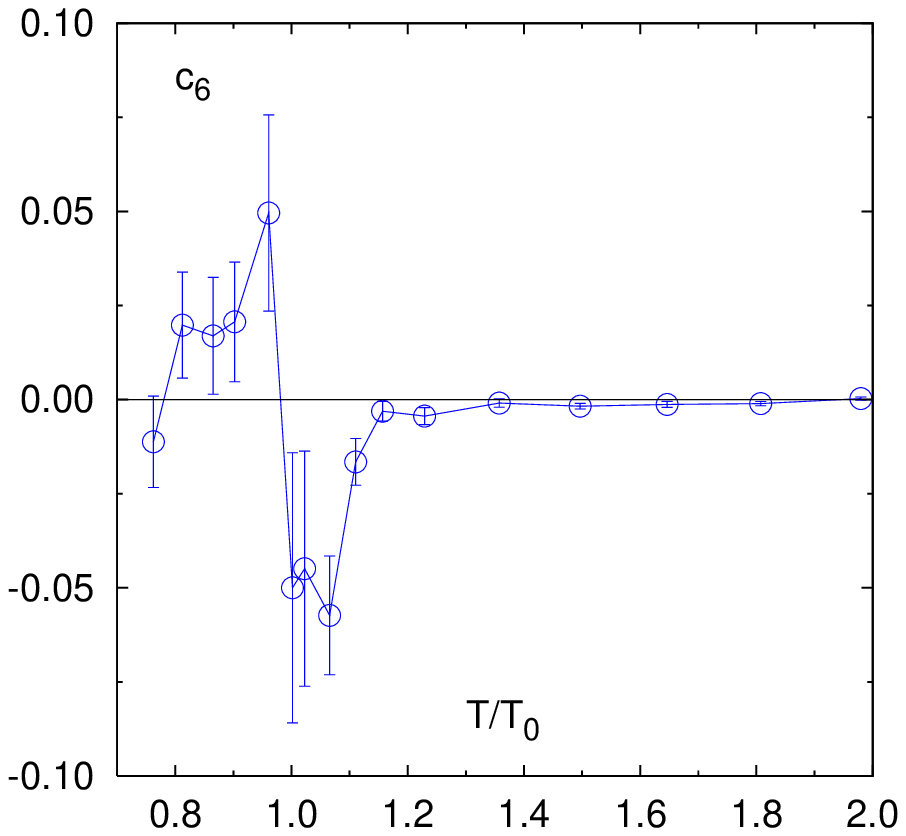}
\end{minipage}
\hfill
\begin{minipage}[b]{0.48\textwidth}
\includegraphics[width=.95\textwidth]{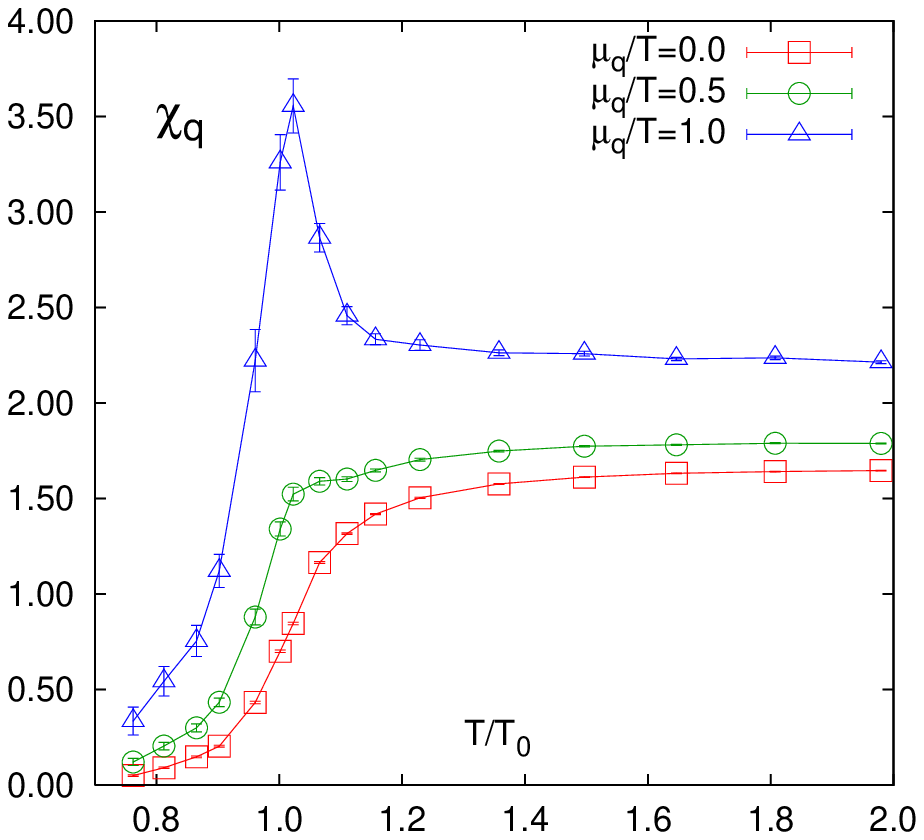}
\end{minipage}
\caption{The Taylor expansion coefficients $c_2$, $c_4$ and $c_6$ of the 
pressure [13] and the quark number fluctuations $\chi_q$ for different values 
of the quark chemical potential.}
\label{fig:c2c4c6}
\end{figure}

Following the transition line into the non-zero chemical potential plane, 
quark number fluctuations $\chi_q$ belong to the most important observables. 
They will diverge at the critical end-point and thus provide an excellent 
signal for the existence and its location on the lattice and eventually may 
be detectable in heavy ion experiments. Hadronic fluctuations can be computed 
from Taylor expansion coefficients of the pressure with respect to the quark 
chemical potential:
\begin{equation}
\frac{p}{T^4}
= \sum_{n=0}^{\infty} c_n(T) 
\left(\frac{\mu_q}{T}\right)^n \quad \mbox{with} \quad 
c_n(T) = 
\left. \frac{1}{n!}\frac{N_t^3}{N_s^3}\frac{\partial {\rm ln} 
Z}{\partial(\hat\mu N_t)^n} \right|_{\hat\mu=0} \quad .
\end{equation}
Due to the particle anti-particle symmetry ($\mu_q \leftrightarrow - \mu_q$) 
all odd coefficients vanish. Thus the first three non-zero coefficients are 
$c_2$, $c_4$, and $c_6$. They have been calculated in the case of two flavors 
of p4-improved staggered fermions, with $m_q/T=0.4$ \cite{Allton:2005gk} and 
are shown in Fig.~\ref{fig:c2c4c6}. 
Note that in the Taylor expansion of the pressure the up and down quark 
chemical potentials have been chosen to be equal. Having calculated the 
coefficients $c_n(T)$ one can construct the quark number density and quark 
number fluctuations
\begin{equation}
\frac{n_q}{T^3}=\sum_{n=2}^{\infty}nc_n(T)\left(\frac{\mu_q}{T}\right)^{n-1}
\quad ; \quad
\frac{\chi_q}{T^2}=\sum_{n=2}^{\infty}n(n-1)c_n(T)
\left(\frac{\mu_q}{T}\right)^{n-2} \quad .
\label{Eq:chiq}
\end{equation}
Also shown in Fig.~\ref{fig:c2c4c6} are the quark number fluctuations for 
various values of the chemical potential. It is interesting to see that at 
$\mu_q=0$, the fluctuations $\chi_q$ show a rapid but monotonic increase at 
the transition temperature, whereas a cusp is developing at $T_c(\mu_q)$ for 
$\mu_q>0$. This is a clear sign for approaching the critical end-point.

\begin{figure}
\begin{center}
\begin{minipage}[b]{0.48\textwidth}
\includegraphics[bb= 100 430 4096 4096,width=.9\textwidth]{qno_combo_new3.ps}
\end{minipage}
\hfill
\begin{minipage}[b]{0.48\textwidth}
\includegraphics[width=.99\textwidth]{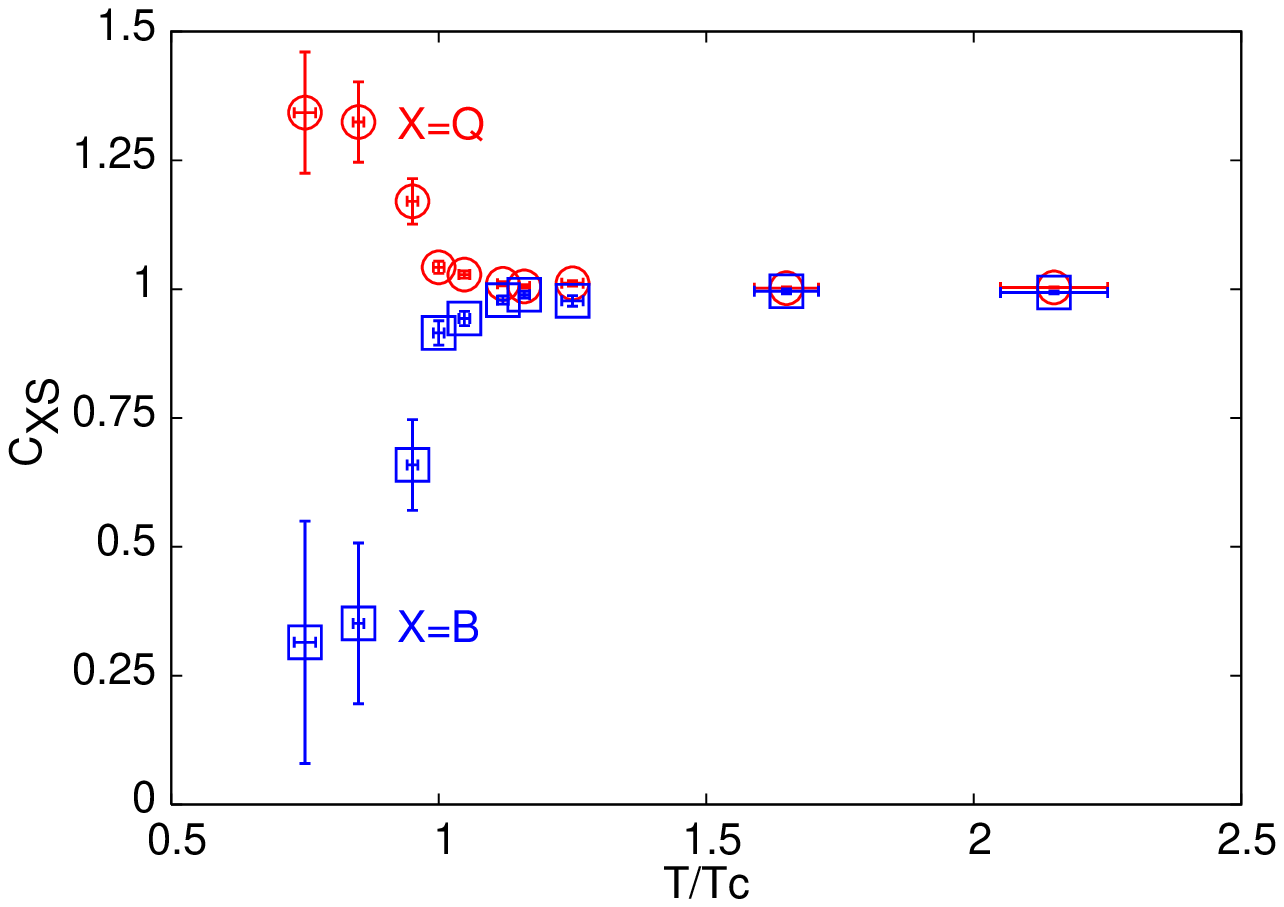}
\end{minipage}
\end{center}
\caption{Different hadronic fluctuations [30] and the correlation between 
strangeness and baryon number, electric charge fluctuations respectively 
[31]. \label{fig:IIYYBB}}
\end{figure}

In the case of 2-flavor QCD, the quark number susceptibility is directly 
proportional to the baryon number fluctuation. Having two light quarks and 
one heavier strange quark, the situation is not that simple anymore. To match
the situations realized in heavy ion collisions, one still wants to expand the 
pressure in terms of $\mu_q=\mu_u=\mu_d$, keeping the strange quark chemical 
potential zero ($\mu_s=0$). However, in order to analyze fluctuations of 
conserved quantum numbers, it appears to be more appropriate to perform a 
basis change going from the space of quark number fluctuations
\begin{equation}
\chi_{\alpha,\beta}=\left< n_\alpha n_\beta \right> - \left< n_\alpha \right> 
\left< n_\beta \right> \propto \frac{T}{V}\frac{\partial^2 {\rm log}Z}{
\partial \mu_\alpha \partial \mu_\beta}\quad,
\end{equation}
where $\alpha,\beta\in\{u,d,s\}$ to the space of hadronic fluctuations, 
indicated  by $\alpha,\beta\in\{I_3,Y,B\}$ with $I_3$ being the third 
component of the isospin, $Y$ being the hypercharge and $B$ the baryon number. 
In Fig.~\ref{fig:IIYYBB} the diagonal susceptibilities $\chi_{I_3,I_3}$, 
$\chi_{Y,Y}$ and $\chi_{B,B}$ as well as the off-diagonal susceptibility 
$\chi_{Y,B}$ are shown
\cite{Bernard:2004je}. 
They have been measured by the MILC Collaboration with 2+1 flavor of Asqtad 
fermions on a $12^3\times 6$ lattice. The light quark mass is $m_q=0.2m_s$ 
where $m_s$ is the physical strange quark mass. The curves have been 
normalized such that the continuum Stefan-Boltzmann value is 0.5 for all of 
them. Qualitatively they show the same behavior as the diagonal quark number 
fluctuations, only the off-diagonal susceptibility $\chi_{Y,B}$ shows a cusp 
already at $\mu=0$. However, up to a minus sign also the off-diagonal quark 
susceptibility $\chi_{u,d}$ shows a cusp at $\mu=0$.

Furthermore, also fluctuations of other conserved quantities such as
electric charge $Q$ have been computed. It is especially interesting to 
analyze the correlations between different conserved charges. Also shown 
in Fig.~\ref{fig:IIYYBB} are the correlations 
\begin{equation}
C_{X,S}=\frac{\left<XS\right>-\left<X\right>\left<S\right>}{\left<S^2\right>
-\left<X\right>^2} \quad,
\end{equation}
where $S$ is the strangeness and the operator $X$ is either the electric 
charge ($X=Q$) or the baryon number ($X=B$) \cite{Gavai:2005yk}. Such
calculations have been performed on an $16^3\times 4$ lattice, using a 
standard staggered action. The two dynamical light quark masses yield 
$m_\pi/m_\rho=0.3$. The strange quark has been treated in the quenched 
approximation. The results show that above $T_c$ strangeness and electric 
charge or baryon number fluctuate independently. This is consistent with the 
quasi-particle picture of the Quark-Gluon-Plasma (QGP). However, below $T_c$, 
in the hadronic phase fluctuations are clearly correlated. These correlations 
are thus directly related to the degrees of freedom in the QGP. These are also
clearly visible in the calculation of higher order cumulants of fluctuations 
\cite{Ejiri:2005wq}. 

\section{The critical end-point}
\label{sec:EP}
Locating the critical point is one of the most challenging tasks for lattice 
QCD at finite chemical potential. The first attempt to locate the critical 
point used the reweighting method \cite{Fodor:2001pe}. For this calculation,
2+1 flavor of standard staggered fermions have been used at a pion mass of 
about 300 MeV and a kaon mass of about 500 MeV. Lattice sizes have, however, 
been rather small ($4^3\times 4$ - $8^3\times 4$). A critical chemical 
potential of $\mu_B^{crit}=725(35)$ was found. A second calculation 
\cite{Fodor:2004nz}, using again the reweighting method, with physical masses 
($m_\pi=150$ MeV, $m_K=500$ MeV) and somewhat larger volume 
($6^3\times 4$ - $12^3\times 4$), let to $\mu_B^{crit}=360(4)$ MeV.  

\begin{figure}
\begin{center}
\begin{minipage}[b]{0.48\textwidth}
\includegraphics[width=.99\textwidth]{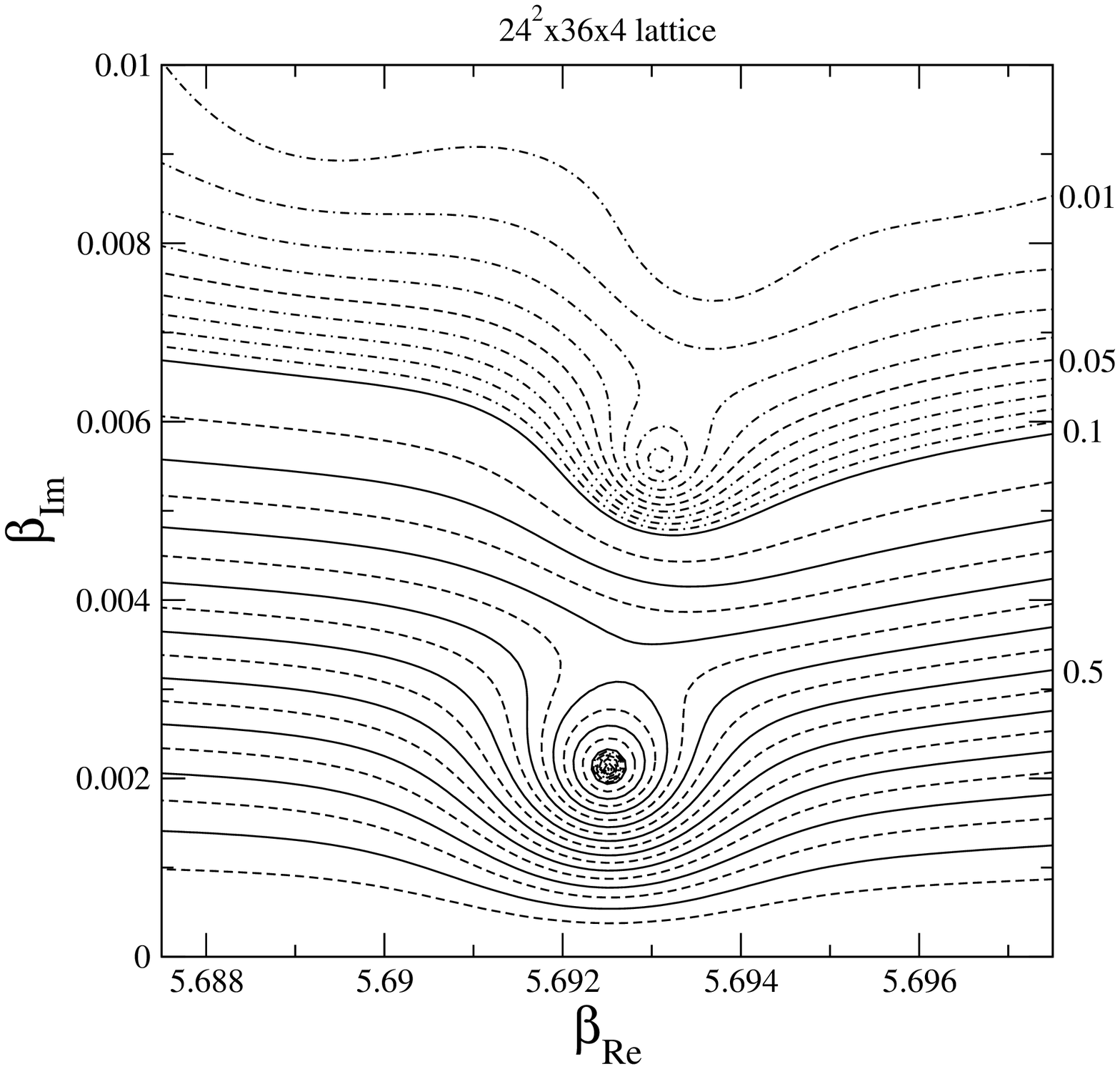}
\end{minipage}
\hfill
\begin{minipage}[b]{0.48\textwidth}
\includegraphics[width=.99\textwidth]{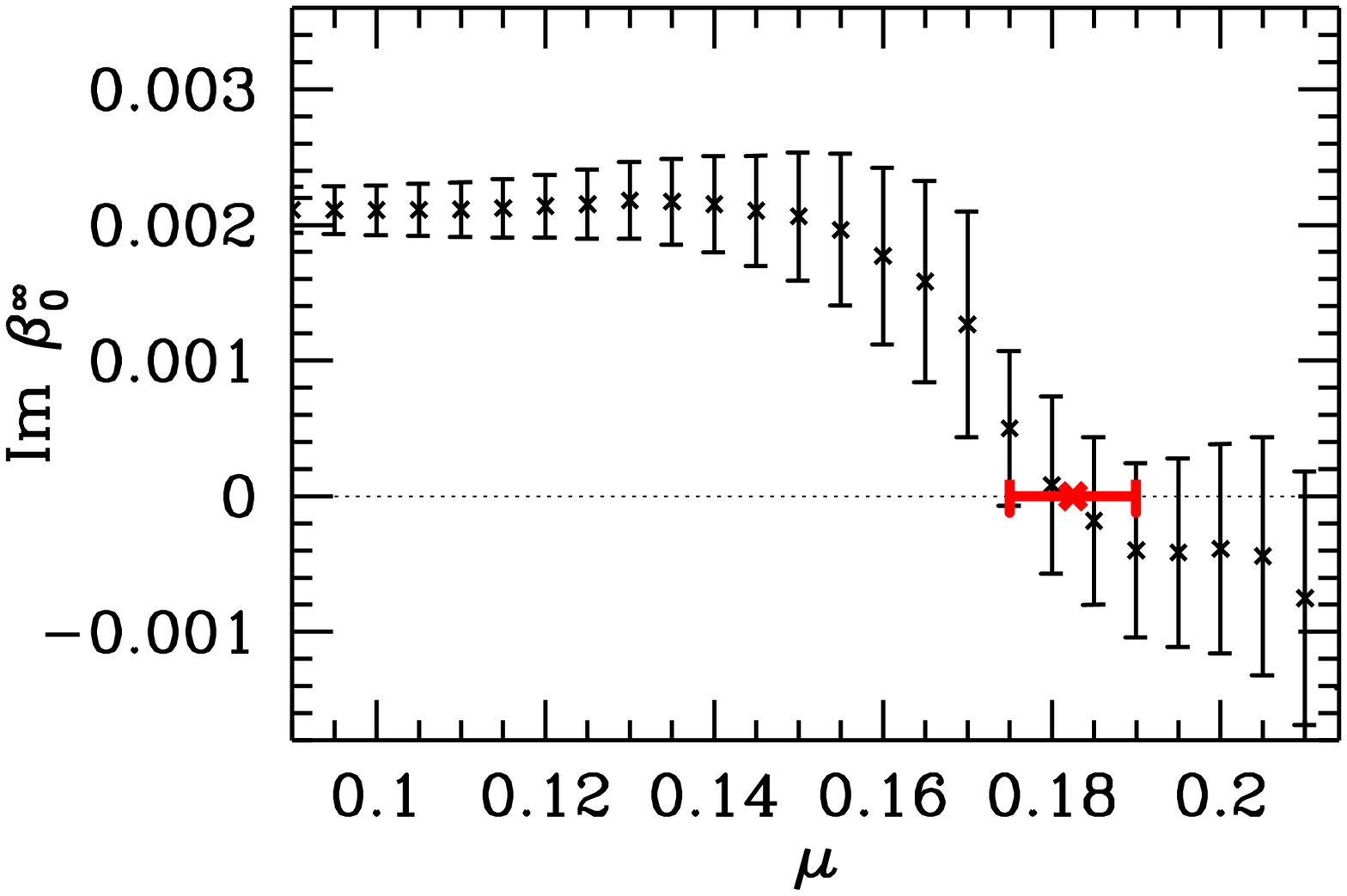}
\end{minipage}
\end{center}
\caption{Lee-Yang Zeroes in the complex $\beta$ plane, in the case of $SU(3)$ 
gauge theory [35] (left) and the distance of the smallest Lee-Yang zero from
the real axis as function of the chemical potential, in the case of full QCD 
[34] (right). \label{fig:leeyang}}
\end{figure}

When using the reweighting method for locating the critical point, the 
minima of the normalized partition function in the complex $\beta$-plane 
(Lee-Yang zeros) have to be determined
\begin{equation}
Z_{\rm norm}\equiv \left| \frac{ Z(\beta_{\rm Re}, \beta_{\rm Im},\mu) }{ 
Z(\beta_{\rm Re}, 0,0)} \right|
=
\left| \left< e^{6i \beta N_t N_s^3 \Delta S_G} e^{i\theta} 
e^{( N_f/4)( {\rm ln det} M(\mu)
- {\rm ln det} M(0) ) }\right>_{(\beta_{\rm Re},0,0)} \right|\quad.
\end{equation}
In $SU(3)$ gauge theory, where we have $e^{i\theta}=1$, this can be done 
with high accuracy, which can be seen in Fig.~\ref{fig:leeyang} 
\cite{Ejiri:2005ts}.
One can even identify a second Lee-Yang zero. In order to locate the critical 
point, one has to take the infinite volume limit and monitor the approach of 
the Lee-Yang zeros on the real axis. When the first Lee-Yang zero touches the 
real axis in the infinite volume limit, a critical point has been reached. In 
Fig.~\ref{fig:leeyang} this is shown for full QCD with physical masses 
\cite{Fodor:2004nz}. In QCD with non-zero chemical potential the analysis of 
Lee-Yang zeros is, however, subtle \cite{Ejiri:2005ts}. For large volumes and 
chemical potentials the phase factor of the determinant $e^{i\theta}$ will 
force the Lee-Yang zero onto the real axis, which might lead to an 
underestimation of the critical point.

Another difficulty with the reweighting method at finite chemical potential 
has been pointed out in \cite{Golterman:2006rw}. It was noted, that taking the 
fourth (or square) root of the determinant (which is necessary in order to 
simulate 2 or 1-flavor QCD with staggered fermions; see also \cite{Sharpe}) 
could lead to phase ambiguities. This problem becomes acute when 
$\mu_q>m_\pi/2$.

All of the a above mentioned limitations are, however, irrelevant for the 
location of the critical point with the reweighting method if the critical 
point is located at small values of $\mu_q$.

Using the Taylor expansion coefficients of the pressure, it is also possible 
to estimate the location of the critical point. The convergence radius of the 
expansion is limited by the nearest singularity in the complex chemical 
potential plane. For each fixed temperature, the radius of convergence is 
given by
\begin{equation}
\rho=\lim_{n\to\infty}\rho_n=\lim_{n\to\infty}\sqrt{\left|
\frac{c_n}{c_{n+2}}\right|}\quad.
\end{equation} 
Moreover, the sign of the coefficients $c_n$ gives information about the 
location of the singularity in the complex plane. If all coefficients are 
positive, the singularity is located on the real axis of the complex chemical 
potential plane. If the sign is strictly alternating, the singularity lies on 
the imaginary axis. For a detailed discussion see \cite{Stephanov:2006dn}.

\begin{figure}
\begin{center}
\begin{minipage}[b]{0.48\textwidth}
\includegraphics[width=.9\textwidth]{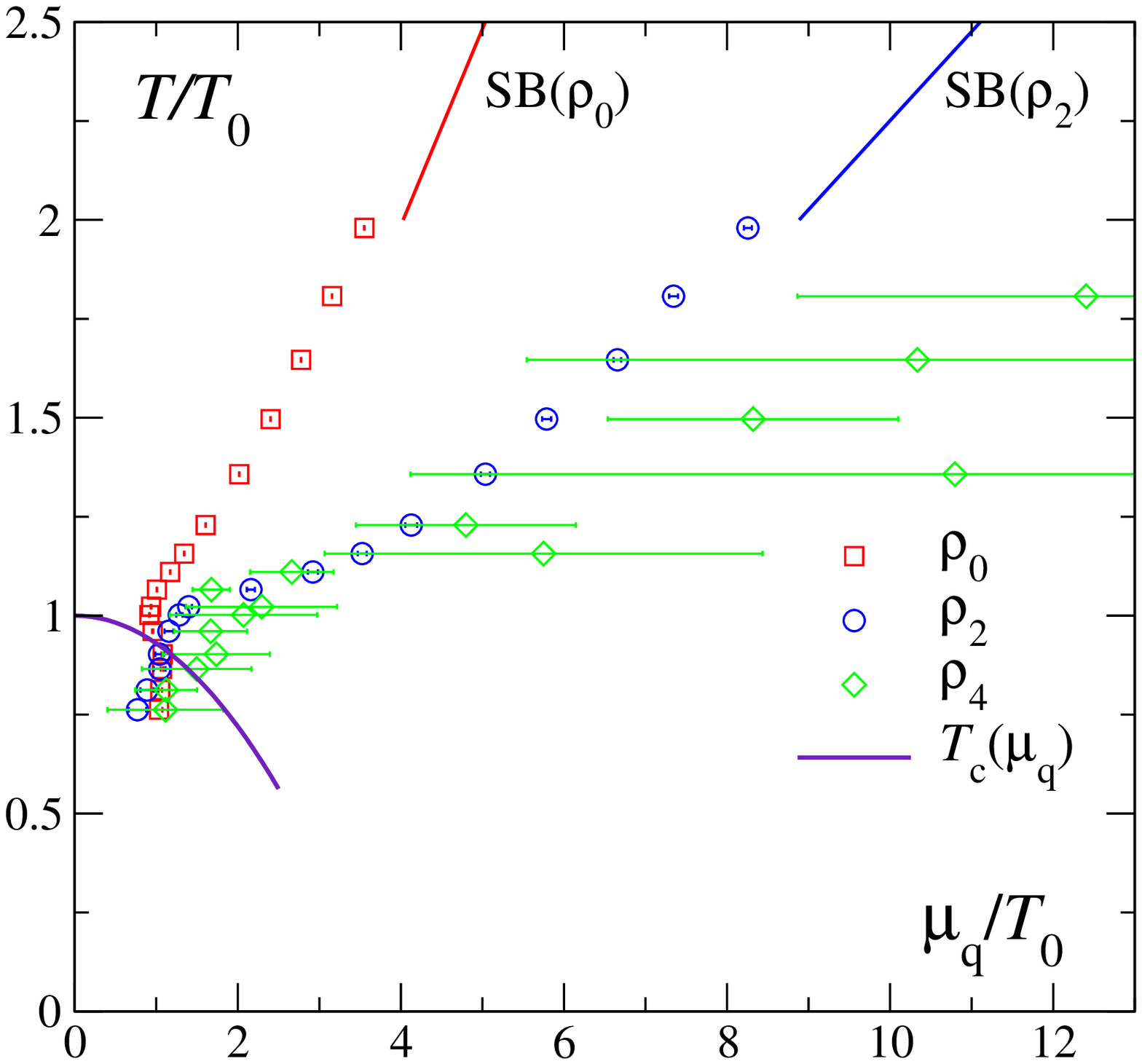}
\end{minipage}
\hfill
\begin{minipage}[b]{0.48\textwidth}
\includegraphics[bb=70 58 320 300, width=.9\textwidth]{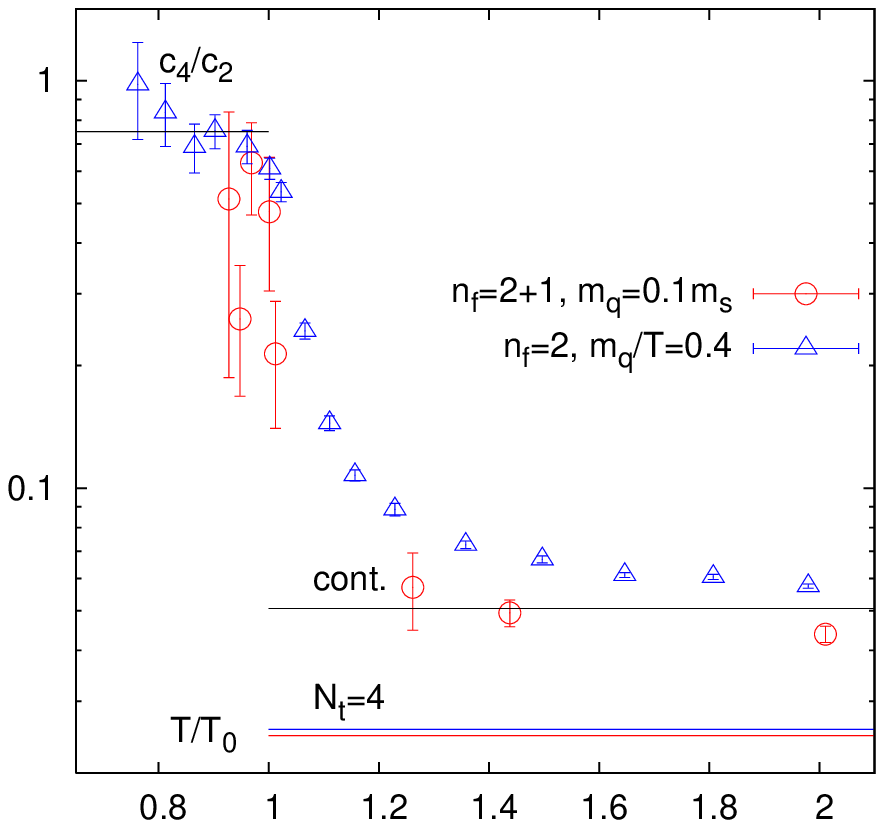}
\end{minipage}
\end{center}
\caption{Estimates of the radius of convergence in the $(T,\mu_q)$-plane 
(left), the ratio $c_4/c_2$ of the expansion coefficients (right). 
\label{fig:rho}}
\end{figure}

Having only a limited number of expansion coefficients, one can only estimate 
$\rho$. The hope is that the convergence of the $\rho_n$ will be fast. Indeed, 
a clustering of the $\rho_n$ is seen in the phase diagram, as shown in 
Fig.~\ref{fig:rho} \cite{Allton:2005gk}.
This calculation, which has been performed with 2 flavors of p4 improved 
fermions and $m_\pi/m_\rho=0.7$, suggests a critical chemical potential of 
$\mu_B^{crit}\approx 500$~MeV. All calculated $\rho_n$ are, however, 
consistent (within statistical error) with the resonance gas model in the 
Boltzmann approximation, where the radius of convergence is infinity. 
 
The authors of \cite{Gavai:2004sd} have estimated the critical chemical 
potential from a Taylor expansion of the quark number susceptibility and 
find $\mu_B^{crit}\approx 180$~MeV. Two flavors of standard staggered fermions 
have been used on lattices up to $24^3\times 4$ and quark mass corresponding 
to $m_\pi/m_\rho=0.3$. The difference between the two estimates 
\cite{Allton:2005gk,Gavai:2004sd} of the critical point is large. We note 
that the second estimate comes from the expansion coefficients of $\chi_q$. 
As can be seen from Eq.~\ref{Eq:chiq} this will result in a smaller $\rho_n$ 
for each fixed $n$. The limit $\lim_{n\to\infty}\rho_n$ is of course the same. 
For finite $n$, however, the estimate of $\mu_B^{crit}\approx 180$ MeV would 
correspond to $\mu_B^{crit}\approx 240$ MeV, when estimating the $\rho_n$ with 
coefficients of the same order from the expansion of the pressure. Nonetheless,
the difference between the two estimates is still striking. The origin 
could be the difference in mass. However, preliminary results from the 
RBC-Bielefeld Collaboration, also shown in Fig.~\ref{fig:rho}, do not indicate
a strong mass dependence in $c_4/c_2=1/\rho_2^2$. 

\begin{figure}
\begin{center}
\begin{minipage}[b]{0.48\textwidth}
\includegraphics[width=.99\textwidth]{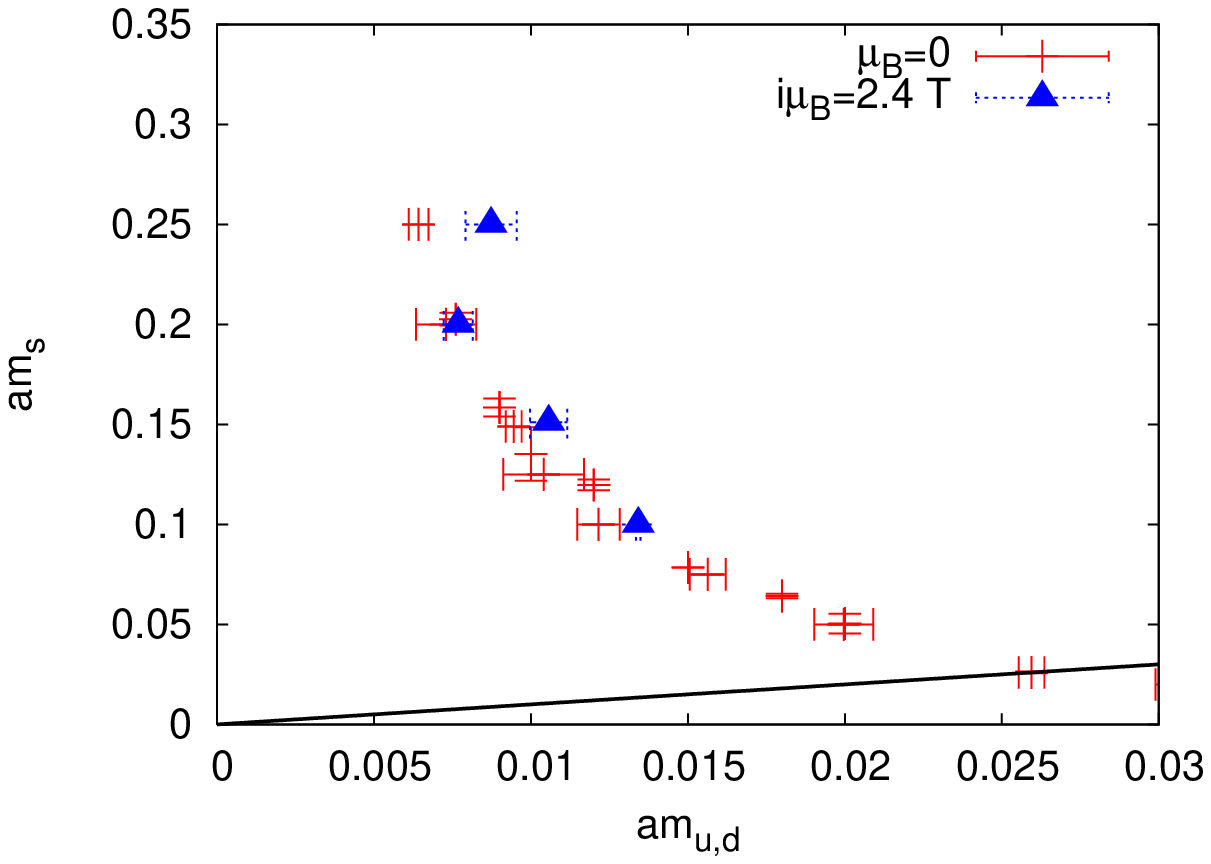}
\end{minipage}
\hfill
\begin{minipage}[b]{0.48\textwidth}
\includegraphics[bb=82 80 388 275, width=.99\textwidth]{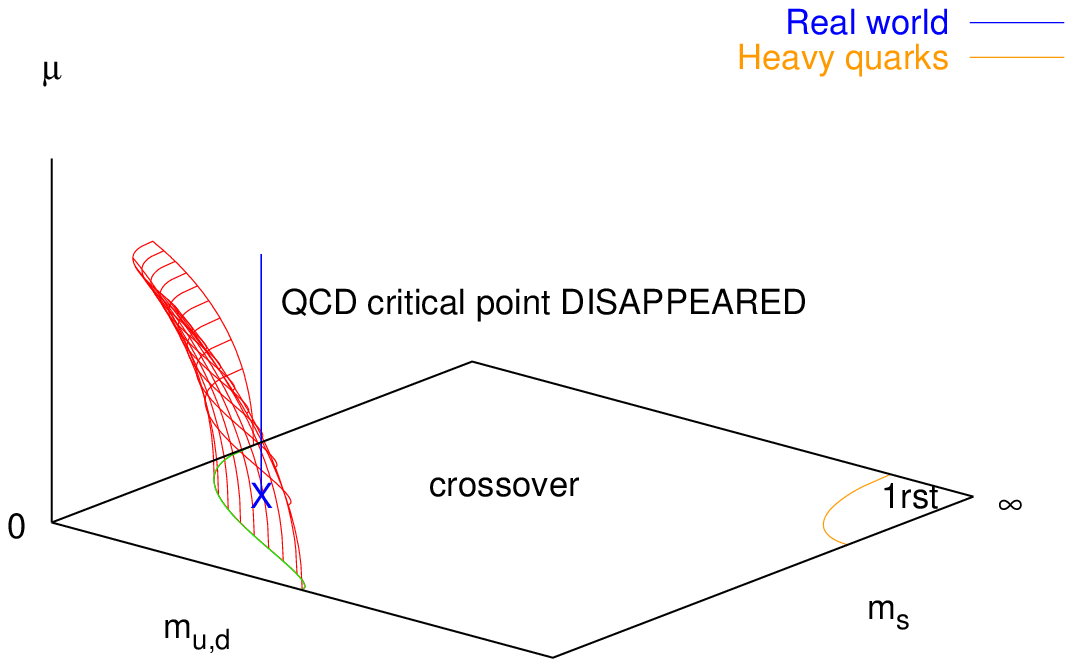}
\end{minipage}
\end{center}
\caption{Line of critical points in the quark mass plane for $\mu_B=0$ and 
$i\mu_B=2.4T$ (left) and sketch of the surface of critical points as found 
in [39] (right). \label{fig:critline}} 
\end{figure}

The critical point can also be studied directly at $\mu_q=0$. This can be done 
by tuning the quark masses carefully to a value where the critical chemical 
potential is $\mu_B^{crit}=0$. In the quark mass plane of two degenerate light 
quark and one strange quark, ($m_{u,d}, m_s$)-plane, a line exists on which 
this condition is fulfilled. Starting from this line, one can define a surface 
of critical points in the 3-dimensional space of ($m_{u,d}, m_s, \mu_q$). On 
one site of the surface, the order of the QCD transition is first order 
(for smaller masses) on the other side the transition is only a smooth 
crossover. The line of critical points at $\mu_q=0$ has been computed for 
standard staggered fermions \cite{deForcrand:2006pv} as shown in 
Fig.~\ref{fig:critline}.
For locating the critical points, the fourth order Binder cumulants of the 
chiral condensate have been calculated. Since the probability distribution 
of the order parameter is universal, also the value of its forth cumulant 
is a renormalization group invariant quantity which furthermore is volume 
independent at the critical point. From a calculation of cumulants at 
imaginary chemical potential the region of first order phase transitions was 
found to shrink as sketched in Fig.~\ref{fig:critline}. This calculation has
been performed with standard staggered fermions on an rather course lattices
($N_t=4$). If this result gets confirmed in the continuum limit it would put
doubts on the existence of a critical point in QCD with physical 
masses. It is interesting to mention, that a shrinking region of first order 
transitions has also been found in the case if isospin chemical 
potential \cite{Sinclair:2006zm}.

\section{Beyond the critical point}
\label{sec:beyond}
\begin{figure}
\begin{center}
\includegraphics[width=.45\textwidth]{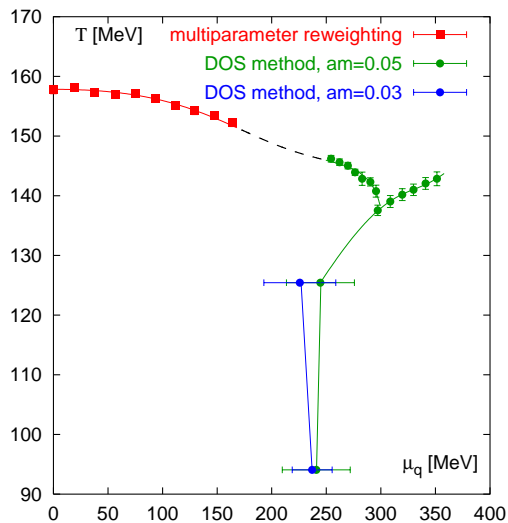}
\caption{Phase diagram from the density of state method [39].\label{fig:dos}}
\end{center}
\end{figure}

Even more challenging than locating the critical point, is the study of the 
physics at high densities and low temperatures. One attempt to do so is a 
calculation using the density of states method \cite{dos}. Using four flavors 
of standard staggered fermions (i.e. taking the root of the determinant is no 
necessary), several simulation points in the $(\beta,\hat\mu)$ plane have been 
chosen to generate phase quenched configurations by employing the method 
proposed in \cite{Kogut:2002zg}. The lattice size has been $6^3\times4$ and 
$6^4$. The quark mass was chosen to be $m/T=0.2$. The generation has been done 
with constrained plaquettes. In oder to do so, the $\delta$-function in 
Eq.~\ref{eq:dos} has been replaced by a sharply peaked Gaussian potential, 
which in practice means that the force term in the HMD-R algorithm had to be 
modified. In the notation used in Sec.~\ref{sec:dos} this would mean $\phi=P$, 
$g=|{\rm det}M|e^{-\beta S_G}$ and $f=e^{i\theta}$. For each simulation
point, several runs have been performed with about 20 different values of the 
plaquette. By calculating the eigenvalues of the reduced matrix the phase of 
the determinant was calculated for each of those runs. By numerically 
calculating the integrals
\begin{equation}
\left<P\right>=\int dx\; x\rho(x)\left<{\rm cos}(\theta)\right>_x \quad
\left<P^2\right>=\int dx\; x^2\rho(x)\left<{\rm cos}(\theta)\right>_x \quad,
\end{equation} 
we recover the grand canonical expectation value of the plaquette and its 
square. Here $\rho(x)$ is the density of states (Eq.~\ref{eq:dos}), which has 
been measured by the integral method, usually used to calculate the pressure. 
The susceptibility of the plaquette is then given by the usual expression 
$\chi_P=\left<P^2\right>-\left<P\right>^2$. From the peak position of the 
plaquette susceptibility the phase diagram was calculated as shown in 
Fig.~\ref{fig:dos}.
The scale was set by the Sommer radius $r_0$, measured on a
$10^3\times 20$ lattice. We find a triple point, where three different phases 
seem to coexist. The phases show different plaquette expectation values. The 
triple point is located around $\mu_q^{\rm tri}\approx 300~\mbox{MeV}$, 
however its temperature ($T^{\rm tri}$) decreases from 
$T^{\rm tri}\approx 148\mbox{MeV}$ on the $6^3\times 4$ lattice to
$T^{\rm tri}\approx 137\mbox{MeV}$ on the $6^4$ lattice. This shift reflects 
the relatively large cut-off effects one faces with standard staggered 
fermions and temporal extents of 4 and 6.

Also shown in Figure~\ref{fig:dos} are points from
simulations with quark mass $m/T=1.2$. The phase boundary turned out to be
--- within our statistical uncertainties --- independent of the mass.

\section*{Acknowledgments}
I would like to thank all members of the RBC-Bielefeld Collaboration for 
helpful discussions and comments. This work has been support by the U.S. 
Department of Energy under contract DE-AC02-98CH1-886.

\end{document}